\documentclass[reprint,amsmath,amssymb,superscriptaddress,citeautoscript,floatfix]{revtex4-1}
\usepackage{graphicx}
\usepackage{color}
\usepackage{dcolumn}
\usepackage{bm}
\usepackage{booktabs}
\DeclareGraphicsExtensions{.png}

\begin{document}

\title{Moment-volume coupling in La(Fe$_{1-x}$Si$_x$)$_{13}$}

\author{M. E. Gruner}
\email{Markus.Gruner@uni-due.de}
\affiliation{Faculty of Physics and Center for Nanointegration Duisburg-Essen (CENIDE), University of Duisburg-Essen, 47048 Duisburg, Germany}
\author{W. Keune}
\affiliation{Faculty of Physics and Center for Nanointegration Duisburg-Essen (CENIDE), University of Duisburg-Essen, 47048 Duisburg, Germany}
\author{J. Landers}
\affiliation{Faculty of Physics and Center for Nanointegration Duisburg-Essen (CENIDE), University of Duisburg-Essen, 47048 Duisburg, Germany}
\author{S. Salamon}
\affiliation{Faculty of Physics and Center for Nanointegration Duisburg-Essen (CENIDE), University of Duisburg-Essen, 47048 Duisburg, Germany}
\author{M. Krautz}
\affiliation{IFW Dresden, P.O.Box 270116, 01171 Dresden, Germany}
\author{J. Zhao}
\affiliation{Advanced Photon Source, Argonne National Laboratory, Argonne, Illinois, 60439, USA}
\author{M. Y. Hu}
\affiliation{Advanced Photon Source, Argonne National Laboratory, Argonne, Illinois, 60439, USA}
\author{T. Toellner}
\affiliation{Advanced Photon Source, Argonne National Laboratory, Argonne, Illinois, 60439, USA}
\author{E. E. Alp}
\affiliation{Advanced Photon Source, Argonne National Laboratory, Argonne, Illinois, 60439, USA}
\author{O. Gutfleisch}
\affiliation{Materials Science, TU Darmstadt, 64287 Darmstadt, Germany}
\author{H. Wende}
\affiliation{Faculty of Physics and Center for Nanointegration Duisburg-Essen (CENIDE), University of Duisburg-Essen, 47048 Duisburg, Germany}

\begin{abstract}
  We investigate the origin
  of the volume change and magnetoelastic interaction observed at the magnetic first-order transition
  in the magnetocaloric system La(Fe$_{1-x}$Si$_x$)$_{13}$ by means of
  first-principles calculations combined with the fixed-spin moment approach.
  We find that the volume of the system varies with the square of the average local Fe moment,
  which is significantly smaller in the spin disordered configurations compared to the ferromagnetic
  ground state. 
  The vibrational density of states
  obtained for a hypothetical ferromagnetic state with artificially reduced spin-moments
  compared to a nuclear inelastic X-ray scattering measurement directly above the phase transition
  reveals that the anomalous softening at the transtion
  essentially depends on the same moment-volume coupling mechanism.
  In the same spirit, the dependence of the
  average local Fe moment on the Si content can account
  for the occurence of first- and second-order transitions in the system.

\end{abstract}
  
\maketitle   

\section{Introduction}
Cooling and refrigeration constitutes a significant part of the world wide energy consumption.
New concepts for
cooling based on solid state materials exploiting magneto-, electro-, elasto- and barocaloric processes
provide a viable alternative to the prevailing gas-compressor schemes
\cite{cn:Moya14NMAT,cn:Faehler12FerroicCooling,cn:Manosa10Baro}.
Based on a century-old idea, research on materials for
refrigeration by adiabatic demagnetization has recently gained significant attention
due to its realistic perspective for application in
room temperature devices \cite{cn:Tishin03,cn:Gschneidner05,cn:Gutfleisch11AM}.
Apart from the classical Gd$_{5}$Si$_{2}$Ge$_{2}$ \cite{cn:Pecharsky97PRL} and $\alpha$-FeRh \cite{cn:Annaorazov92},
which are too expensive for bulk applications, and NiMn-Based Heusler-alloys, which
are still on an exploratory level, La(Fe$_{1-x}$Si$_x$)$_{13}$ \cite{cn:Fujieda02APL,cn:Lyubina10AM}
and Fe$_2$P-based \cite{cn:Dung11AEM,cn:Boeije16CM} compounds are currently the most promising candidates for
practical application (see
\cite{cn:Brueck17SSP,cn:Gutfleisch16PhilTrans,cn:Liu12NMAT,cn:Morrison12PhilMag,cn:Sandeman12Scripta}
for recent reviews).

As a ternary system, LaFe$_{1-x}$Si$_x$ undergoes a ferromagnetic (FM)
to paramagnetic (PM) phase transition with a Curie
temperature $T_{\rm C}$ significantly below room temperature. For application purposes, however, $T_{\rm C}$ can be
shifted to ambient conditions by adding substitutional elements like Co or by loading with
interstitial Hydrogen together with subtitutional Mn
\cite{cn:Fujita03H,cn:Wang03CP,cn:Barcza11IEEE,cn:Krautz14JALCOM}.
For the knowledge-based optimization of these materials, a basic understanding of their intrinsic properties on the
electronic scale is of fundamental importance.
This includes first-principles calculations of the electronic structure of ternary and quaternary compositions
\cite{cn:Fujita03,cn:Wang06JMMM,cn:Han08,cn:Kuzmin07,cn:Boutahar13,cn:Fujita12Scripta,cn:Gercsi14,cn:Gercsi15EPL,cn:Fujita16APLM},
M\"ossbauer spectroscopy for the magnetic properties
\cite{cn:Hamdeh04,cn:Wang06JMMM,cn:Makarov15JPD},
structural properties from neutron diffraction
\cite{cn:Wang03JPCM,cn:Rosca10JALCOM} and even spin wave stiffness or Debye temperature from
measurements of the low temperature specific heat \cite{cn:Lovell16PRB}.
Thus apart from materials engineering such as
optimization of microstructure and thermodynamic properties, a large amount of efforts has been
dedicated to fundamental research. Indeed, it has been revealed early, that the
technologically interesting first-order phase transition
transition observed at low Si content is caused by itinerant electron metamagnetism
\cite{cn:Fujita99,cn:Fujita01}, i.\,e., the competition of
two metastable magnetic states of the Fe atoms. These are associated with a large change in unit cell volume of 
about 1\,\% for a Si content of $x$$\,=\,$1.5 measured at $T_{\rm C}$, which is consequently
accompanied by a large barocaloric effect \cite{cn:Manosa11NCOMMS}.

In addition to improving the intrinsic properties, which are directly relevant
for the magnetocaloric effect, such as
the adiabatic temperature change $\Delta T_{\rm ad}$, the isothermal entropy change $\Delta S$ or
the magnetization change $\Delta M$, another important target for materials optimization is
reducing the hysteresis at the first order transition
\cite{cn:Fujita14APL,cn:Lovell15AEM,cn:Gutfleisch16PhilTrans}.
The large volume change $\Delta V$ is believed to play an important role in this respect,
as it establishes an additional obstacle for the propagation of the interface between the FM and the PM
phase and is thus likely associated with energy dissipation.
$\Delta V$ reduces significantly when the FM-to-PM transition becomes second-order upon increasing the Si
content \cite{cn:Palstra83,cn:Jia06JAP}.
Simulation of the transition behavior of La-Fe-Si based on empirical modelling
suggests that this is related to the strength of the intrinsic magnetoelastic coupling \cite{cn:Basso14IJR},
which is sensitive to the composition. 

First-principles computations of the Fe-projected vibrational density of stated (VDOS) in the FM and
PM state were found to be in excellent agreement with results obtained by
nuclear resonant inelastic X-ray scattering (NRIXS),
and characteristic changes in the VDOS across $T_{\rm C}$ were shown to originate
from the itinerant electron metamagnetism associated with Fe and from
pronounced magneto-elastic softening in the Fe subsystem in the PM
state \cite{cn:Gruner15PRL}. The results demonstrate that significant changes occur in the
Fe-projected VDOS of LaFe$_{11.6}$Si$_{1.4}$ when the sample is heated from the
low-temperature FM state to the high-temperature PM state. In
particular, a distinct phonon mode at a phonon energy of about 27\,meV that
exists in the FM state is completely quenched in the PM
state. The combination of theory and experiment shows that the entropy change at the phase transition
is connected to cooperative contributions from magnetism, lattice and electronic structure, which all
have the same sign. In turn, previous modelling approaches and thermodynamic analysis of experimental data,
achieve good agreement based on the assumption of a competition between magnetic
and lattice degrees of freedom \cite{cn:Piazzi16JMMM,cn:Ranke05,cn:Jia06JAP}.
This implies that the common decomposition of the entropy
change into essentially independent degrees of freedom is insufficient and
neglects unknown, but potentially important coupling terms involving appropriate combinations of the
state variables. A necessary prerequisite for an appropriate analytic
formulation of such terms is a detailed understanding of the coupling mechanisms between the microscopic
degrees of freedom.

The purpose of the present contribution is to give a detailed account on the interplay of electronic, magnetic and vibrational
degrees of freedom in La-Fe-Si from parameter-free first-principles calculations.
We will present the site-resolved magnetic moments of Fe
as a function of the total magnetization and Si content and relate this to features in the electronic structure,
which were recently made responsible for the pronounced changes in vibrational properties at $T_{\rm C}$.
We will prove by comparing the vibrational density of states from first-principles calculations
involving a hypothetical FM state with
constrained magnetization and corresponding NRIXS data obtained directly above $T_{\rm C}$
that spin disorder is not directly involved in the magnetoelastic coupling mechanism.
Finally, our work will demonstrate that -- for the ternary system --
the magneto-elastic coupling
mechanisms leading to improved intrisic magnetocaloric properties are also responsible for
the volume change and thus potentially increase
thermal hysteresis at the same time.

\section{Methodological details}
\begin{figure}[tb]%
  \centering
\includegraphics*[width=0.8\linewidth]{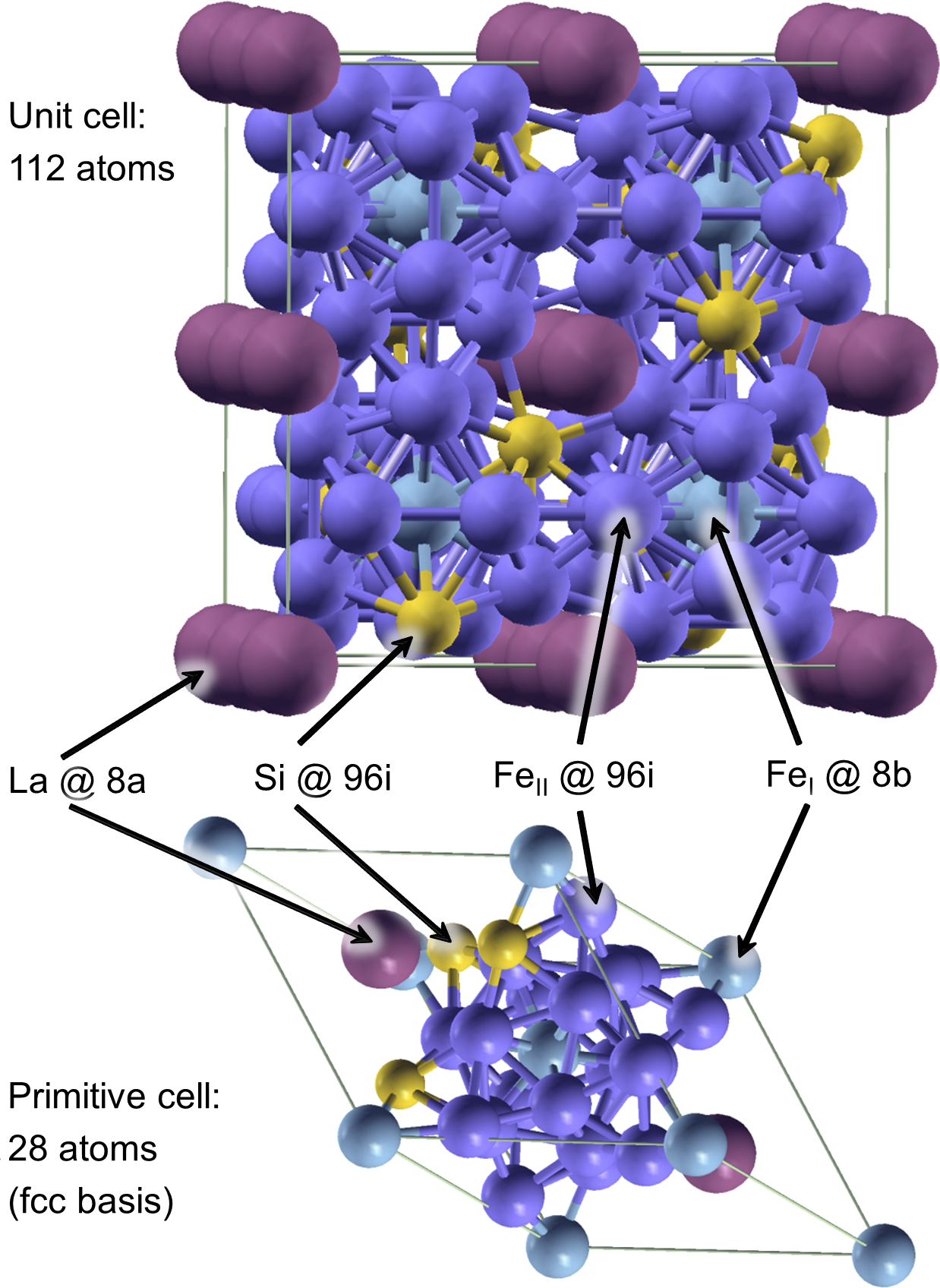}
\caption{%
  Lattice structure of cubic LaFe$_{11.5}$Si$_{1.5}$ with space group 226 (Fm$\overline{3}$c).
  La (purple spheres) resides on the 8a Wyckoff position, Fe$_{\rm I}$ (bright blue) on 8b, while
  Fe$_{\rm II}$ (dark blue) and Si (yellow) share the 96i sites. In the calculation,
  a pseudo-disordered configuration was used,
  where Si resides on specific 96i site such that rhombohedral symmetry (space group 146) is retained.
  Top panel: Unit cell with 112 atoms and Cartesian basis as used in the calculations.
  Bottom: Primitive cell with two formula units (28 atoms) and fcc basis vectors.}
  \label{fig:Cell}
\end{figure}
\subsection{Primitive cell representation}
The unit cell of LaFe$_{13-x}$Si$_x$ contains 112 atoms
arranged in a cubic structure with $Fm\overline{3}c$ symmetry (space group 226)
and Cartesian basis vectors, as shown on the upper panel of Fig.\ \ref{fig:Cell}.
The La atoms occupy the (8a) Wyckoff positions, while Si is found in the (96i) sites.
Two chemically different types of Fe exist. Fe$_{\rm I}$ is found on the (8b) positions, while
Fe$_{\rm II}$ shares with Si the (96i) sites \cite{cn:Chang03JoPD,cn:Hamdeh04,cn:Rosca10JALCOM} .
The 112 atom unit cell is large enough for a random placement of Si on
the (96i) sites to generate sufficiently accurate statistics
for practical purposes. Consequently, we used such configurations for benchmark purposes
and for testing the dependence of magnetism on chemical composition.
However, to accurately assess the lattice dynamics, which is a central aspect of the present work,
we need a multitude of single calculations according to the lack of configurational symmetry.
In an extended large supercell this easily exceeds current computational capabilities.

Previous approaches have off-stoichiometric 1:13 phases either by placing Si on the (8b) sites
\cite{cn:Kuzmin07} -- which is incorrect from the crystallography point of view but retains the symmetry --
or use an analytic description of disorder within
the virtual crystal or the coherent potential approximation \cite{cn:Fujita12Scripta,cn:Fujita16APLM}.
All of these approaches take full advantage of the high
cubic symmetry with three inequivalent lattice sites.
In the latter cases, a relaxation of the structure with
respect to interatomic forces is not feasible anymore, which effectively
prevents the calculation of a vibrational properties, while in the former case,
the vibrational properties of Si, which is now residing in the center of the icosahedral
cage instead of being at the edges, may not be reproduced accurately anymore.

There is, however, another possibility to represent the NaZn$_{13}$ 
structure in a compact 28 atom primitive cell with a fcc basis.
If one appropriately chooses 3 of the 24 original (96i) sites to be
occupied with Si, one can retain at least
rhombohedral symmetry (space group 146, {\em R3}),
see the bottom of Fig.\ \ref{fig:Cell}.
This leads to a nominal composition of LaFe$_{11.5}$Si$_{1.5}$ with 12
inequivalent lattice sites, i.\,e., two La plus two Fe$_{\rm I}$, each
singly occupied and one Si plus seven Fe$_{\rm II}$ sites, being three-fold
occupied, each. The disadvantage is that one treats a pseudo-ordered
compound, where artificial correlations in the elemental distributions
exist which may alter the result. It is important to mention that
despite the rhombohedral symmetry of the cell no significant
deviation in the angles between the lattice vectors were found. Thus,
all configurations considered here can be
described by cubic fcc-type lattice vectors.

\subsection{Computational details}
The parameter-free first-principles calculations were carried out in the
framework of the density
functional theory (DFT) \cite{cn:Hohenberg64}, which yields, in principle,
the ground state properties of a system as total
energy, electronic structure, magnetic moments and
interatomic forces from the knowledge of atomic position and the
number of electrons only.  In practice, additional approximations are
necessary which involve technical parameters, which are given below.

We made use of the Vienna Ab-initio Simulation
Package (VASP) \cite{cn:VASP1} which is efficiently parallelized
allowing for the treatment of large supercells on
massively parallel computer hardware \cite{cn:Gruner09Review}.
VASP achieves an excellent compromise between speed and accuracy by
describing the wave functions of the valence electrons
using a plane wave basis set while taking advantage of the
projector augmented wave (PAW) approach \cite{cn:VASP2},
which takes care of the
interaction with the core electrons.
For the accurate description of structural properties of
ferrous systems, the use of
the generalized gradient approximation (GGA)
for the representation of the exchange-correlation
functional is mandatory.
As in our previous {\em ab initio} study for ordered Fe$_3$Pd \cite{cn:Gruner11PRB}, we
used the PW91 formulation by
Perdew and Wang \cite{cn:Perdew91,cn:Perdew96a}
in connection with
the spin interpolation
formula of Vosko, Wilk and Nusair \cite{cn:Vosko80}.
We calculated in the scalar-relativistic approximation
with a collinear spin setup.
A detailed comparison of the PW91 PAW potentials with PBE potentials
\cite{cn:Perdew96} and full potential calculations 
with both GGA functionals
can be found in Ref.\ \cite{cn:Gruner11PRB} for the case of Fe$_3$Pd.
We used a valence electron configuration of 5s$^2$5p$^6$5d$^1$6s$^2$
for La, $3d^74s^1$ for Fe and $3s^23p^2$ for Si. The corresponding
cutoff for the electronic structure calculations was 
$E_{\rm cut}$$\,=\,$$335$\,eV.
For the structural optimizations (cell parameter and atomic positions)
in the 28 atom cell we used a $k$-mesh of 7$\times$7$\times$7
Monkhorst-Pack
grid  which yields 60 $k$-points in the irreducible Brillouin zone (IBZ)
and a finite
temperature smearing according to Methfessel and
Paxton \cite{cn:Methfessel89}
with a broadening of $\sigma$$\,=\,$0.1\,eV.
For the structural optimization in the disordered 112 atom cell, we restricted
to a 2$\times$2$\times$2 Monkhorst-pack k-mesh. 
The self-consistency cycle was stopped when the difference in energy
between two consecutive cycles fell below $10^{-7}$\,eV. The
optimized structures had residual forces of less than
$5\,\times\,10^{-3}\,$eV/\AA{}.
To obtain an accurate account on the electronic structure, the
calculations in the 28 atom cell
were followed by a single step without relaxation carried out with
a Monkhorst-Pack k-mesh of up to 13$\times$13$\times$13 (371 points in the
IBZ) and the tetrahedron method with Bloechl
corrections \cite{cn:Bloechl94}.
Forces and pressure were checked to remain small. 
For the presentation in the main article, the electronic density of
states (DOS) generated from these data were again convoluted with a Gaussian
broadening with $\sigma$$\,=\,$0.1\,eV in order to highlight
the most important features.

\begin{figure*}[tb]%
  \centering
  \includegraphics*[width=.9\textwidth]{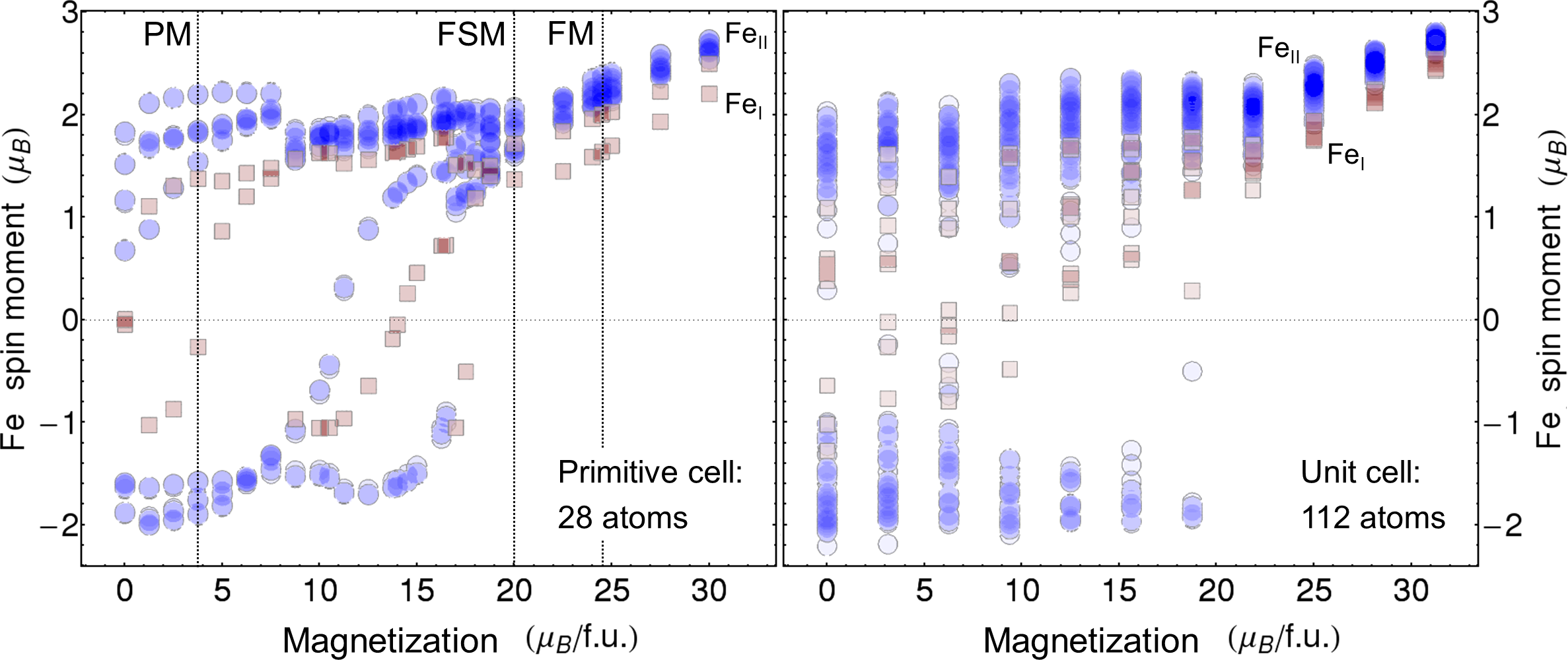}%
  \caption{%
Site-resolved atomic spin moments
as a function of constrained magnetization
per formula unit obtained with the fixed spin moment procedure described in the text
for the 28 atom primitive cell (left) and 112 atom unit cell (right).
Only Fe-moments are shown. Moments on the Fe$_{\rm I}$ sites (Wyckoff position 8b) are shown as
brown squares, while blue circles refer to moment on the Fe$_{\rm II}$ sites. When the symbols overlap the color becomes more saturated, indicating that several moments are close to each other.
Again, the vertical lines denote the three magnetic states, FM, PM and FSM, discussed in the text.
}
\label{fig:Moment}
\end{figure*}

\subsection{Calculating the vibrational density of states}
The theoretical determination of the vibrational density of states (VDOS)
is based on the so-called direct or force-constant approach
\cite{cn:Kresse95,cn:Gonze97,cn:Parlinski97}.
The dynamical matrix
is obtained from the Hellman-Feynman forces resulting from
small atomic displacements of selected atoms, one at a time.
The number of displaced atoms and their direction depends
on the crystal symmetry space group. For the FM and PM configuration a
minimum of 28 individual displacements are required to construct the
force-constant matrix, which is Fourier-transformed to obtain
the dynamical matrix.
Its eigenvalues establish the phonon dispersion relations and the
vibrational density of states.
The corresponding analysis was carried out with the {\sc PHON} code
written by Dario Alf\`e{} \cite{cn:Alfe09PHON}, which also generated
the necessary displacements.
Similar first-principles calculations for Ni$_2$MnGa
have reached excellent agreement between theory and experiment
\cite{cn:Uijttewaal09,cn:Ener12,cn:Dutta16PRL}.

For the DFT part, we used the minimum supercell of 2$\times$2$\times$2
primitive cells, containing 224 atoms in total, which is necessary
for extrapolation of the dispersion in the full BZ. 14 $k$-points
in the IBZ proved sufficient for obtaining
accurate forces.
Brillouin zone integration was again carried out with
the Methfessel-Paxton finite
temperature integration scheme (smearing parameter
$\sigma$$\,=\,$0.1\,eV) \cite{cn:Methfessel89}.
We made sure that residual forces remained
below $10^{-3}$\,eV/\AA{} in the unperturbed supercell.
Then forces were calculated
for 28 independent displacements of $0.02\,$\AA{} in size each, which
were taken in positive and negative directions, in order to improve
accuracy further. Thus, in the end 112 individual super-cell calculations
were carried out to get the FM and PM dynamical matrices.
To obtain the vibrational density of states, $g(E)$,
we used a mesh of 61$\times$61$\times$61 points in
the reciprocal phonon $q$-space and convoluted the results with an
additional Gaussian broadening
(smearing parameter of $\sigma$$\,=\,$1\,meV) corresponding to the
experimental resolution.

\begin{figure}[t]%
  \centering
\includegraphics*[width=0.95\linewidth]{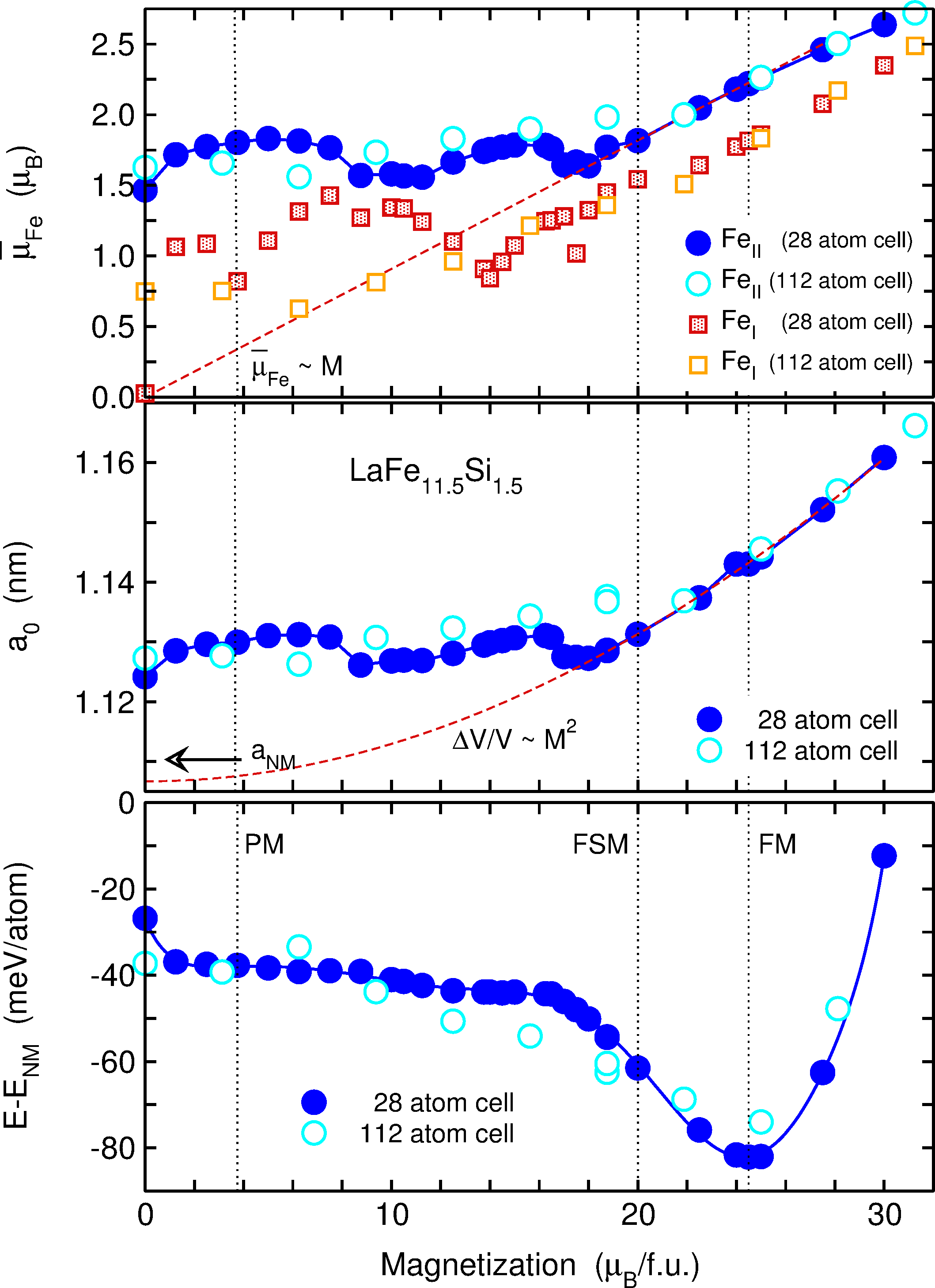}
\caption{%
  Total energy $E$ (bottom panel),
  optimized lattice parameter $a_0$ (center panel) and average magnetic moment
  per site $\overline{\mu}_{\rm Fe}$ (top panel) as a function of constrained magnetization
  per formula unit obtained with the fixed spin moment procedure described in the text.
  The energy is specified relative to the non-spinpolarized (NM) state.
  the horizontal lines denote the ferromagnetic (FM) ground state at $M$$\,=\,$$24.5\,\mu_{\rm B}$/f.u.
  the spin-configurations with $M$$\,=\,$$3.75\,\mu_{\rm B}$/f.u. used as model for the paramagnetic state
  (PM) and the ferromagnetic configuration with an artificially reduced moment of
  $M$$\,=\,$$20\,\mu_{\rm B}$/f.u. (FSM). The red dashed line in the center panel refers to a variation of
  the volume-magnetostriction $\omega$$\,=\,$$\Delta V/V$ which is expected to be
  proportional to the square of the magnetic moment.
  The arrow indicates the equilibrium volume of a non-spinpolarized (NM) state.
  In the upper panel, the blue circles refer to the average magnetic moment at the Fe$_{\rm II}$ site
  (open: 112 unit atom cell, filled: 28 atom primitive cell), while the reddish squares describe the atomic
  moments averaged over the Fe$_{\rm I}$ sites. For comparison, the red dashed line illustrates
  a coherent linear variation of the average moments $\overline{\mu}_{\rm Fe}$
  with the magnetization per cell $M$.
}
\label{fig:EVM}
\end{figure}
\subsection{Vibrational density of states from experiment}
 In order to measure the vibrational (phonon) density of states
 (VDOS), g(E), of the Fe subsystem in LaFe$_{11.6}$Si$_{1.4}$, $^{57}$Fe NRIXS \cite{cn:Seto95PRL,cn:Sturhahn95PRL,cn:Chumakov95EPL}
 was performed at the Sector-3 beamline at the Advanced Photon Source,
 Argonne National Laboratory.
 We used a powder sample with nominal composition LaFe$_{11.6}$Si$_{1.4}$
 and with Fe enriched to 10\,\% in the isotope $^{57}$Fe to enhance the NRIXS
 signal. The sample was produced by arc-melting in pure Ar atmosphere
 and was subsequently annealed at 1323\,K for 7 days in an Ar-filled
 quartz tube followed by quenching in water. The powder was made by
 crushing the same material (annealed and quenched ingot) as studied
 in our former work \cite{cn:Gruner15PRL}. The sample was characterized by X-ray
 diffraction and M\"ossbauer back-scattering \cite{cn:Gruner15PRL}, showing that 89\,\% of
 the Fe atoms in the sample are in the La(Fe,Si)$_{13}$ phase and only 11\,\%
 are in the bcc Fe secondary phase.  Further details of the sample
 preparation and sample characterization are described in
 Refs. \onlinecite{cn:Landers,cn:Gruner15PRL,cn:Liu11Acta,cn:Krautz14JALCOM}.
 Magnetic characterization of the powder material was
 performed using the vibrating sample magnetometer
 (VSM) option of a Quantum Design PPMS DynaCool. Our magnetization-versus-temperature measurements (not shown)
 revealed a first-order FM-to-PM phase transition at $T_{\rm C}$$\,\approx\,$190\,K under
 an applied magnetic field $\mu_0\,H$$\,\approx\,$0.7\,T, with a small hysteresis $\Delta T$ of
 $\approx$\,3 K.
 $^{57}$Fe NRIXS spectra were taken with the sample (LaFe$_{11.6}$Si$_{1.4}$
 powder embedded in epoxy resin) exposed again to an external field $\mu_0\,H$$\,\approx\,$0.7\,T,
 produced by a pair of small permanent magnets surrounding the sample. The
 incident x-ray energy was around $E_0$$\,=\,$14.4125\,keV, being the nuclear
 resonance energy of the $^{57}$Fe nucleus. After passing through a
 high-resolution crystal monochromator, the x-ray beam had an energy
 bandwidth of 1\,meV \cite{cn:Toellner00}. A toroidal mirror was used to collimate the
 monochromatized beam onto the sample surface. A closed-cycle cryostat
 for sample cooling was employed for the experiment. An avalanche
 photodiode (APD) detector was placed right outside the dome-shaped Be
 window of the cryostat
 to collect delayed nuclear decay radiation after phonon-assisted 
 nuclear resonant excitation as the NRIXS signal.
 The Fe-specific VDOS was extracted from the NRIXS data using the PHOENIX
 program \cite{cn:Sturhahn00,cn:Sturhahn04}. The actual LaFe$_{11.6}$Si$_{1.4}$-sample temperature was
 determined from the rule of ‘detailed balance’, which is intrinsic
 to NRIXS spectra and uses the fact that in the measured phonon
 sidebands the ratio of the contribution from phonon annihilation,
 $S(-E)$, and phonon creation, $S(+E)$, at phonon energy $|E|$ is equal to
 the Boltzmann factor, $\exp(-|E|/k_{\rm B}T)$ \cite{cn:Sturhahn04}.

\section{Results}
\subsection{Calculations with constrained magnetization: Site-resolved moments}
While it is straight-forward to set-up the ferromagnetic (FM)
representative of the off-stoichiometric compound, it is not obvious, how to generate
a suitable stable paramagnetic (PM) configuration.
We derived this from a systematic procedure starting from the
fully relaxed ferromagnetic structure.
In this procedure, we employed the
fixed spin moment method (FSM)
\cite{cn:Schwarz86} to constrain the total
magnetization $M$ of the entire 28 atom cell in small subsequent
steps, starting from the ferromagnetic ground state configuration.
In each step, the cell parameters and atomic positions were
allowed to adapt freely after the change of the 
magnetization $M$ of the entire cell.
Similar calculations were also carried out for the 112 atom unit cell with random placement
of Si on the (96i) sites. 
The site-resolved configurations of magnetic moments of Fe 
according to this procedure are shown in Fig.\ \ref{fig:Moment} for both cell sizes.
They essentially exhibit a similar pattern, but particular features appear smeared out
in the 112 atom cell, which provides a more realistic representation of the disorder of
Fe$_{\rm II}$ and Si on the (96i) sites.
In the FM ground state ($M$$\,=\,$$24.5\,\mu_{\rm B}$/f.u.),
the Fe$_{\rm II}$ sites exhibit slightly larger magnetic moments compared to
Fe$_{\rm I}$ with a small but noticeable variation in magnitude.
La and Si (not shown) exhibit small induced moments, with
antiferromagnetic coupling to the surrounding Fe.
Above $M$$\,\approx\,$$20\,\mu_{\rm B}$/f.u.,
the atomic moments change coherently and proportionally to the magnetization of the cell.
Below this value, the distribution of moments broadens increasingly and
individual spins spontaneously reverse their direction --
at first the Fe$_{\rm I}$ spins, with further decreasing $M$ also Fe$_{\rm II}$.
Nevertheless, the average $Fe_{\rm II}$ moments retain a more or less stable absolute value
around $\approx$$\,1.7\ldots 1.9\,\mu_{\rm B}$ and the constraint $M$ is predominantly
tuned by the spin configuration. The average spin magnetic moment per Fe$_{\rm II}$ atom
$\overline{\mu}_{\rm Fe}$ shown in the upper panel of Fig.\ \ref{fig:EVM}
clearly confirms this change in trend around $M$$\,\approx\,$$20\,\mu_{\rm B}$/f.u.
for both simulation cells.

Still, there are some Fe-spins found with a significantly reduced
magnetic moment. These appear predominantly, when 
the magnetization constraint forces specific sites to flip their moment
when two magnetic configurations are competing.
Such low spin Fe moments have also been observed in Fe$_2$P-type systems \cite{cn:Boeije16CM}.
The presence of low magnetic moment states of Fe is heavily discussed
as the origin of moment-volume anomalies, such as the Invar effect,
observed in several ferrous alloys \cite{cn:Wassermann90}.
The most prominent example is probably fcc $Fe_{65}Ni_{35}$ where at ambient conditions
thermal expansion is practically compensated by volume-magnetostriction. 
Indeed, strong thermal expansion anomalies are observed in LaFe$_{13-x}$Si$_{x}$,
showing up in terms of a large, discontinuous change of the lattice parameters
at the isostructural FM-PM transition and a reduced thermal expansion in the FM phase
\cite{cn:Chang03JoPD,cn:Jia06JAP}.
In Fig.\ \ref{fig:Moment}, low moment states of Fe with (absolute) spin magnetic moments smaller
than 1\,$\mu_{\rm B}$ are found predominantly on the Fe$_{\rm I}$ site,
despite their comparatively small presence in the compound. This allows us to speculate, that
in the spirit of previous work on ferrous Invar alloys
\cite{cn:Kaspar81,cn:Moruzzi90,cn:Entel93,cn:Heike99}, the high
coordination of the Fe$_{\rm I}$ atoms in the center of the corner-sharing icosahedra
may foster the instability of the magnetic moment.
The implication of the change in Fe spin moment on the
lattice parameter will be discussed in the next section.

Constraining the exchange splitting of the cell in general
requires the artificial decoupling of the Fermi-level, i.e., the chemical potential of the electrons,
with respect to the
spin-up and spin-down channel. This is equivalent to applying different magnetic fields to
the two spin channels, which is, of course, not possible in experiment.
Therefore, realistic spin configurations should have
the same chemical potential in both spin channels.

For the 28 atom cell, we encounter for the
spin configuration with $M$$\,=\,$$3.75\,\mu_{\rm B}$/f.u that both Fermi-levels nearly
coincide. Consequently, it was possible to regain this spin configuration 
without FSM constraint and prove its metastable nature in this way.
In turn, this configuration, denoted by the leftmost dotted vertical line in Fig.\ \ref{fig:Moment}
was chosen as a representative for the PM state in our calculations. In contrast to a truly
disordered configuration our PM is rather a kind of uncompensated antiferromagnet. It retains
rhombohedral symmetry even when taking into account the magnetic configurations, which is beneficial
for calculating the vibrational properties of the PM in Ref.\ \cite{cn:Gruner15PRL}.
On the other hand, the comparison to the 112 atom cell, where all symmetry is completely removed
due random placement of Si on the (96i) sites, shows an analogous magnetic behavior.

All calculations were carried out with collinear spin configuration and quenched moments
may indicate an instability of a specific magnetic structure, which cannot be lifted
in a collinear setup.
Since the magnetic exchange constants for mapping the system onto a Heisenberg model have not been
determined, yet, nothing is known about possible competing ferro- and antiferromagnetic
interactions, which may induce non-collinear spin structures.
Such a situation has been proposed for the fcc Fe-based Invar systems, earlier \cite{cn:Schilfgaarde99}.
Therefore,
we carried out additional non-collinear calculation for the PM configuration, which finally
confirmed that the collinear PM configuration
does not decay into another magnetic structure and is thus at least metastable. This also applies
the quenched Fe$_{\rm I}$ moment, which is present in our PM configuration.

Apart from the FM ground state with $M$$\,=\,$$24.5\,\mu_{\rm B}$/f.u.
(rightmost dotted vertical line  in Fig.\ \ref{fig:Moment}, we will discuss in detail
the (artificial) homogeneous
ferromagnet with the constrained magnetization of
$M$$\,=\,$$20.0\,\mu_{\rm B}$/f.u., denoted by FSM (central dotted vertical
line  in Fig.\ \ref{fig:Moment}) which is
situated right next to the configurations with flipped spins, but exhibits very similar
absolute atomic moments. This artificial state will help to understand the impact of the absolute magnitude
of the magnetic moments on the elastic properties ({\em moment-volume} coupling)
as opposed to the disordered spin configuration ({\em spin-lattice} coupling).

\subsection{Magnetoelastic properties and moment-volume coupling}
  The 28 atom cell shows a FM ground state minimum at
  $M$$\,=\,$$24.5\,\mu_{\rm B}$/f.u., while the energy increases
  strongly with any enforced variation of $M$ in both directions (cf.\ Fig.\ \ref{fig:EVM}).
  For $M$$\,<\,$18$\,\mu_{\rm B}$/f.u. the slope decreases and the
  total energy reaches a plateau-like behavior,
  only with minor variation down to essentially $M$$\,=\,$0. The energy between the FM ground
  state and a hypothetical nonmagnetic state $E_{\rm NM}$ without spin-polarization on any site
  amounts to sizable 80\,meV/atom. Within the 128 atom unit cell, we obtain a similar picture.
  However, the variation of the energy from the FM ground state to the PM state appears smoother
  and the clear change in slope at intermediate $M$ becomes smeared out. This may be a consequence of
  the disordered arrangement, whereas the rhombohedral symmetry constraint present
  in the 28 atom cell enforces that three Fe$_{\rm II}$ sites flip
  their direction simultaneously, which is an artificial constraint not present in real disordered
  compounds. Taking into account the differences in the choice of the
  simulation cell and the technical setup, these results are consistent,
  while there is also reasonable agreement with previous calculations
  of ordered LaFe$_{12}$Si$_1$ using the full potential localized orbital (FPLO) code \cite{cn:Kuzmin07}
  and VASP \cite{cn:Gercsi14}
  and disordered, off-stoichiometric LaFe$_{11.5}$Si$_{1.5}$ obtained with the
  KKR-CPA method \cite{cn:Fujita12Scripta,cn:Fujita16APLM}.
  
  The energies refer  for each magnetization to the respective
  equilibrium lattice parameter $a_0$, shown in the center panel of Fig.\ \ref{fig:EVM} which
  was obtained in an additional optimization procedure. As for the total energy,
  there is around  $M$$\,\approx 18\ldots20\,\mu_{\rm B}$/f.u.
  a change in the dependence of $a_0$ on $M$, which is much steeper for the larger
  magnetizations. This coincides with the linear change of the site-resolved Fe-moments with $M$
  (see again Fig.\ \ref{fig:Moment} and the upper panel of Fig.\ \ref{fig:EVM}),
  which is enforced where spin-flips cannot accommodate for a mismatch. The considerably flatter part
  at lower $M$ is also reflected in the average atomic moments. From the comparison of the upper and
  center panel of Fig.\ \ref{fig:EVM} we recognize a close correlation between the average moment on the
  Fe$_{\rm II}$ sites and $a_0$, for both the 28 atom primitive cell and the 128 atom unit cell.

  From the theory of itinerant magnetism, one expects the volume magnetostriction, i.\,e.,
  $\Delta V/V$ to vary proportionally to the square of the magnetic moment $M$,
  which is indeed fully obeyed by this compound. 
  This is demonstrated by the red dashed curve in the center panel of Fig.\ \ref{fig:EVM}.
  The relation provides an excellent fit for larger $M$, where spin-flips are absent, while it
  deviates significantly for smaller $M$, where the average atomic spin moments $\overline{\mu_{\rm Fe}}$
  deviate from the proportionality to $M$. In turn,
  comparing the extrapolation of this curve to $M$$\,=\,$$0$
  once again agrees well with the optimized lattice constant of a non-spinpolarized configuration
  with zero spin-moment on every site.
  The correlation is not as pronounced for the Fe$_{\rm I}$ sites. This may be in part ascribed to the fact
  that per formula unit there are about ten times more Fe$_{\rm II}$ compared to Fe$_{\rm I}$.  
  An important detail in this respect may be the positioning of the
  Fe$_{\rm I}$ within the icosahedral Fe$_{\rm II}$ cages, which could take up part of the
  chemical pressure and are thus relevant for the expansion behavior. 

\begin{figure}[t]%
  \centering
\includegraphics*[width=0.95\linewidth]{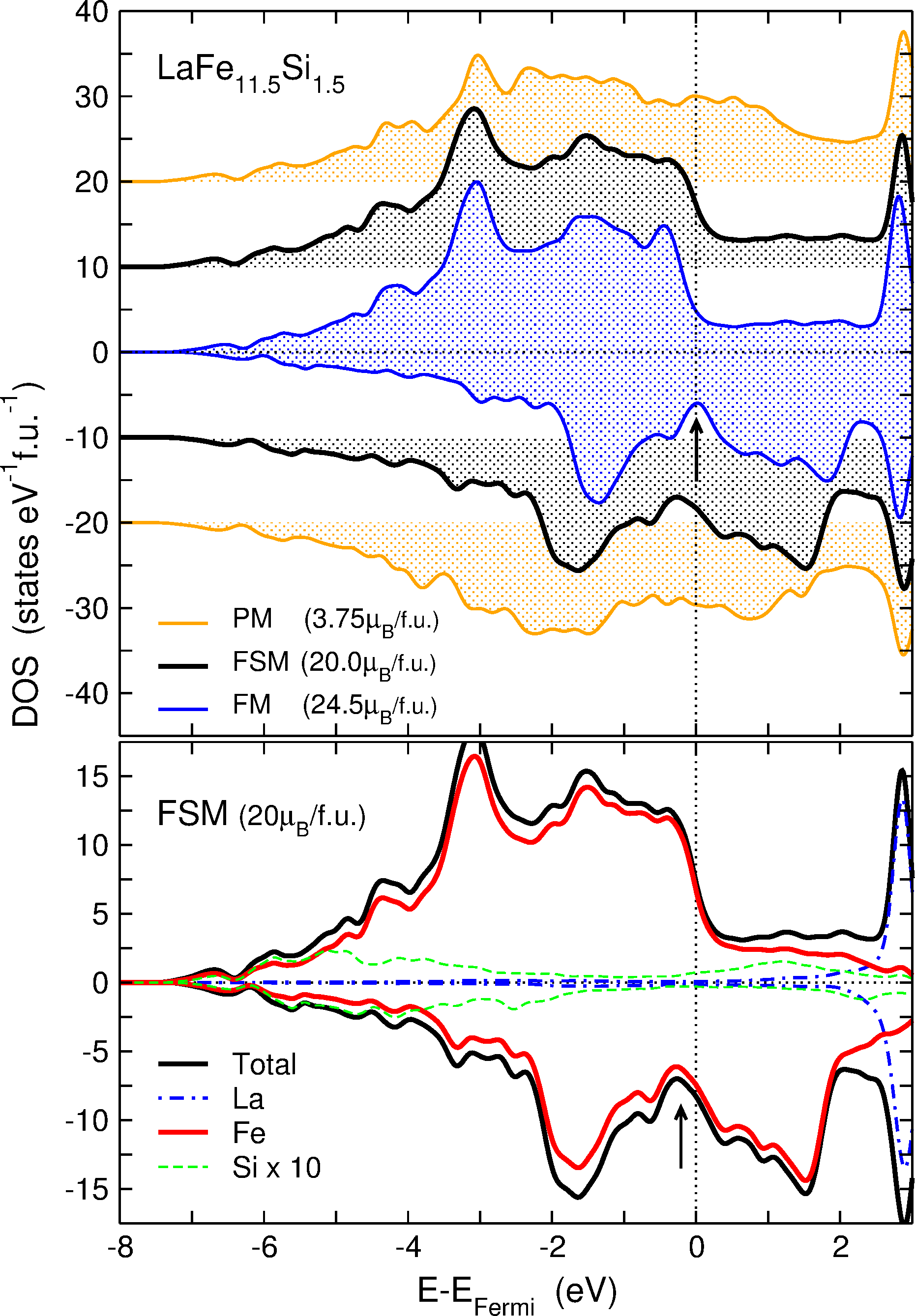}
\caption{%
  Spin-polarized electronic density of states (DOS) of LaFe$_{11.5}$Si$_{1.5}$ for different magnetic states.
  Positive values refer to the majority spin channel, negative values to the minority spin channel.
  The upper panel compares the total VDOS of the FM configurations at 24.5$\,\mu_{\rm B}$/f.u.
  (thin blue line), the approximated PM
  at 3.75$\,\mu_{\rm B}$/f.u. (thin orange line) and the FSM constrained stated at 20.0$\,\mu_{\rm B}$/f.u.
  (thick black line).
  To improve visibility, the FSM and the PM DOS are shifted by
  a constant value.
  The arrow denotes the pronounced dip in the FM minority spin channel
  which is responsible for the adiabatic electron phonon coupling.
  The lower panel shows the element resolved DOS of the FSM state. The features in the total DOS
  are dominated by the Fe-contribution (red line). The dip in the minority spin DOS (arrow)
  has moved away from the Fermi level (vertical dotted line). The data for the FM and PM state were taken from
  Ref. \protect\cite{cn:Gruner15PRL}.
}
\label{fig:DOS}
\end{figure}
  \subsection{Adiabatic electron phonon coupling and spin disorder}\label{sec:vibrational}
  Based on the above findings, we can now clarify the detailed role of spin disorder for itinerant metamagnetism
  in La-Fe-Si. Following our earlier argument \cite{cn:Gruner15PRL}, the presence of 
  neighboring Fe-sites with inverted spin channels changes the hybridization of d-electrons
  in both spin channels.  This effectively reduces the average exchange splitting on the Fe-sites and
  increases the density of states at the Fermi level \cite{cn:Gruner15PRL}, as demonstrated in the upper panel of
  Fig.\ \ref{fig:DOS}. Of particular importance is the pronounced minimum in the minority spin density of states (DOS)
  appearing in the FM phase directly at the Fermi-level $E_{\rm Fermi}$ (marked by the arrow), which has
  essentially disappeared  in the PM electronic structure. A pronounced minimum in the DOS directly at $E_{\rm Fermi}$
  is a stabilizing factor for any phase, since it inhibits changes in the electronic structure, which involve a
  redistribution of states across the Fermi-level. These can for instance occur due to a change in band width,
  which is sensitive to the interatomic spacing (i.e., changes in lattice parameters or isotropic volume change)
  and the exchange splitting, i.\,e., changes in the magnetic moment per atom.
  This makes it plausible to ascribe the increased magnetic moment per Fe atom combined
  with a reduced thermal expansion to the presence of the comparatively sharp minimum at $E_{\rm Fermi}$. In turn,
  forcing the system out of this minimum requires crossing an free energy barrier. This is reflected in the
  lower panel of Fig.\ \ref{fig:EVM} in terms of a comparatively steep increase in energy at either side of
  the FM ground state, in particular for the 28 atom cell. 
  The competition between two metastable magnetic configurations, separated by a free energy barrier gives rise
  to the itinerant electron metamagnetism pointed out earlier and is also reason for the first-order nature
  of the transition. This, again, is subject to thermal hysteresis, which depends on the shape of the free energy barrier.
  
  In the PM phase, the availability of states at the Fermi level is significantly increased, in particular for the orbitals
  in the minority channel, which are spatially less contracted than their majority spin counterparts.
  This allows the electronic
  subsystems to accommodate perturbations arising from changes of the ionic positions more efficiently, 
  which finally gives rise to an anomalous
  softening of vibrational modes in the PM phase,
  previously dubbed adiabatic electron phonon coupling \cite{cn:Delaire08PRL,cn:Delaire11PNAS,cn:Munoz11PRL}.
  The flat energy profile in the low magnetization part of Fig.\ \ref{fig:EVM} suggests
  that a stable configuration with a decreased average Fe$_{\rm II}$ moment might possibly
  be stabilized thermodynamically by the large spin entropy of the paramagnet, but not
  for intermediate magnetic disorder.

  Further stabilization of the PM phase
  comes from the increase in lattice entropy arising from the lattice softening.
  The underlying mechanism requires a change in Fe-moment, which is sufficient to increase the DOS at
  $E_{\rm Fermi}$, but not necessarily spin disorder.
  This can be seen directly by looking at a hypothetical intermediate ferromagnetic configuration, where
  the spin moments are all parallel, but artificially reduced in size due to the FSM constraint.
  We found a particularly interesting configuration at a total moment of $20\,\mu_{\rm B}/$f.u. (FSM)
  stabilized by the fixed spin moment procedure.
  Here, the local moment per Fe is nearly reduced to its PM value, without any
  spin disorder in the configuration. The reduced exchange splitting shifts the minimum in the minority spin DOS
  to lower energies, which in consequence increases the DOS at the Fermi level.

\begin{figure}[t]%
  \centering
\includegraphics*[width=0.95\linewidth]{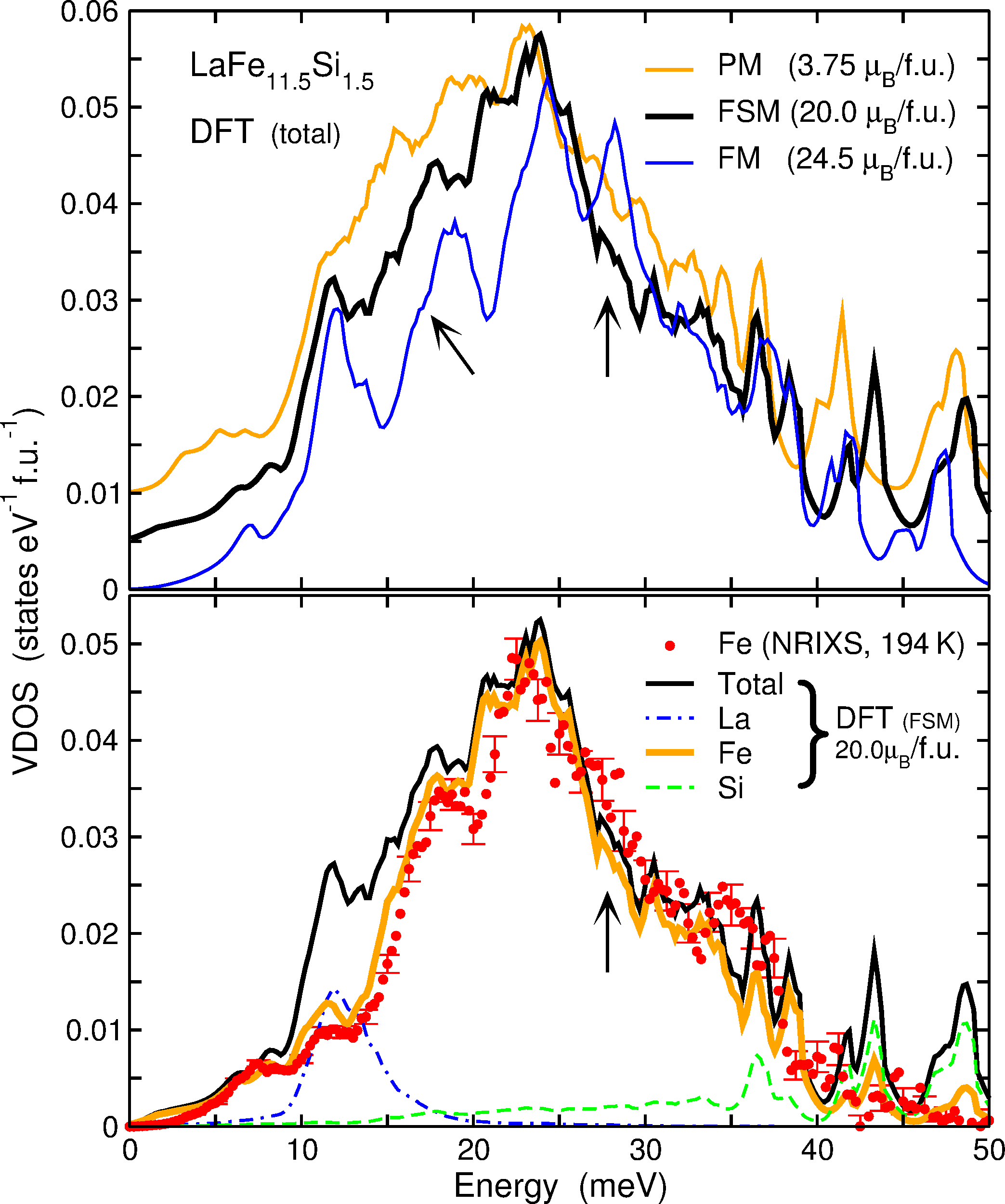}
\caption{%
  Vibrational density of states (VDOS) of LaFe$_{11.5}$Si$_{1.5}$ for different magnetic states.
  The total VDOS of the FM configurations at 24.5$\,\mu_{\rm B}$/f.u. (thin blue line), the approximated PM
  at 3.75$\,\mu_{\rm B}$/f.u. (thin orange line) and the FSM constrained stated at 20.0$\,\mu_{\rm B}$/f.u.
  (thick black line) are compared in the upper panel. Despite lacking magnetic disorder, the FSM VDOS already exhibits major
  characteristics of the PM state, such as the disappearance of the strong peak at 28\,meV and the
  red shift at lower frequencies arising from the partial Fe-contribution. Lower panel: Total and
  element resolved
  VDOS of the FSM state (Total: black, Fe: orange, La: blue, Si: green)
  in comparison to the experimental partial VDOS of Fe (red circles with error-bars) obtained from
  NRIXS at $T$$\,=\,$$194\,$K, directly above the phase transition.
  The theoretical data for the FM and PM state were taken from
  Ref. \protect\cite{cn:Gruner15PRL}.
}
\label{fig:VDOS}
\end{figure}

It is now straight-forward to calculate the vibrational density of states (VDOS) for
  the FSM configuration as we did for the FM and PM configurations in our earlier work \cite{cn:Gruner15PRL}.
  As the DOS is characteristic for the interactions in the electronic subsystem, the VDOS provides
  a fingerprint for the lattice dynamics arising from the forces between the atoms.
  The adiabatic electron phonon coupling is now responsible for the changes in the electronic structure at
  the FM-PM transition leading to characteristic changes in the VDOS (compare the blue and the orange curves
  in the upper panel of Fig. \ref{fig:VDOS}). The most prominent are firstly
  the red-shift of the PM VDOS, which softens the lattice. This occurs despite a rather
  strong decrease in lattice constant, which according to Gr\"uneisen theory is expected to harden the phonons.
  Secondly, we find the pronounced peak at 27\,meV in the FM phase to disappear in the PM phase.
  The FSM VDOS shares the most important features with the PM VDOS, the absence of the 27\,meV peak
  as well as the red-shift, albeit to a slightly lesser extent as the magnetically disordered case.
  Other features of the FM VDOS, such as the marked peaks at 12\,meV and 19\,meV are still present, but
  by far not as pronounced. The thermodynamic properties associated with the FSM VDOS will be discussed
  elsewhere \cite{cn:Landers}.
  
  It was shown earlier,
  that the calculated Fe contributions of both, FM and PM VDOS
  match excellently the experimental partial Fe-VDOS measured by NRIXS
  at low (FM) and ambient temperatures (PM) \cite{cn:Gruner15PRL}.
  Therefore, we compare the Fe-partial VDOS with NRIXS data obtained
  at 194\,K in a magnetic field of 0.7\,T. Under these conditions, the magnetic phase transition takes
  place at $T$$\,=\,$190\,K. Again, the overall agreement between theory and experiment is very good --
  in particular with respect to the anomalous red-shift, which is only slightly overestimated by the calculations.
  The experimental Fe VDOS in the PM state at 194 K, as
  exhibited in Fig.\ \ref{fig:VDOS} , is in excellent agreement with the Fe-projected
  VDOS previously obtained at 220\,K, as shown in Fig. S8 in the
  supporting information of Ref.\ \onlinecite{cn:Gruner15PRL}, where
  the peak at $\approx\,$27\,meV  has disappeared entirely.
  At 194\,K, we see due to the proximity to the phase transition still some faint remainders of the 27\,meV peak.
  This might be related to compositional inhomogeneities in the sample, which have the consequence
  that small fractions of the sample may still be in the FM phase.
  Further details of a NRIXS study on LaFe$_{11.6}$Si$_{1.4}$ near the
  transition temperature will be published elsewhere in near future \cite{cn:Landers}.
  
  The close agreement between experiment and theory shows, that the changes in the electronic structure
  related to the decrease of the Fe-moments alone are
  sufficient to explain the observed changes in the VDOS at the FM-PM transition -- spin disorder does not take
  part directly in the coupling between electronic degrees of freedom and magnetism.
  Based on these findings, we can conclude that the anomalous vibrational behavior of La-Fe-Si
  is rather related to the {\em moment-volume} interaction than {\em magnon-lattice} or
  {\em spin-lattice} coupling. The fact that anomalous softening and volume change have the same origin
  has important consequences for the optimization of this system:
  Attacking hysteresis losses by reducing the volume change at the metamagnetic transition will necessarily
  diminish the change in lattice entropy, and thus reduce the magnetocaloric
  performance of the material intrinsically.

\subsection{Electronic properties and magnetic moments as a function of the Si content}
Silicon plays a decisive role for the formation of the 1:13 compound. The cubic NaZn$_{13}$ prototype structure
is not stable for pure LaFe$_{13}$. For its formation, substitution of at least 7-10\,at.-\% Fe by Si is required
\cite{cn:Niitsu12,cn:Han08}. Adding further Si drives the transition from first to second-order.
This leads to a gradual increase of the transition temperature and a decrease of the volume change and
hysteresis width at the FM-PM transition \cite{cn:Liu12Scripta}.
On the other hand, it is observed experimentally, that the
isothermal entropy change  decreases, which is a measure of intrinsic magnetocaloric
performance \cite{cn:Palstra83,cn:Jia06JAP}.
At still larger Si-content, $x$$\,\geq\,$$2.6$ a transition to a tetragonal structure was reported \cite{cn:Han08}.

\begin{figure}[t]%
  \centering
\includegraphics*[width=0.95\linewidth]{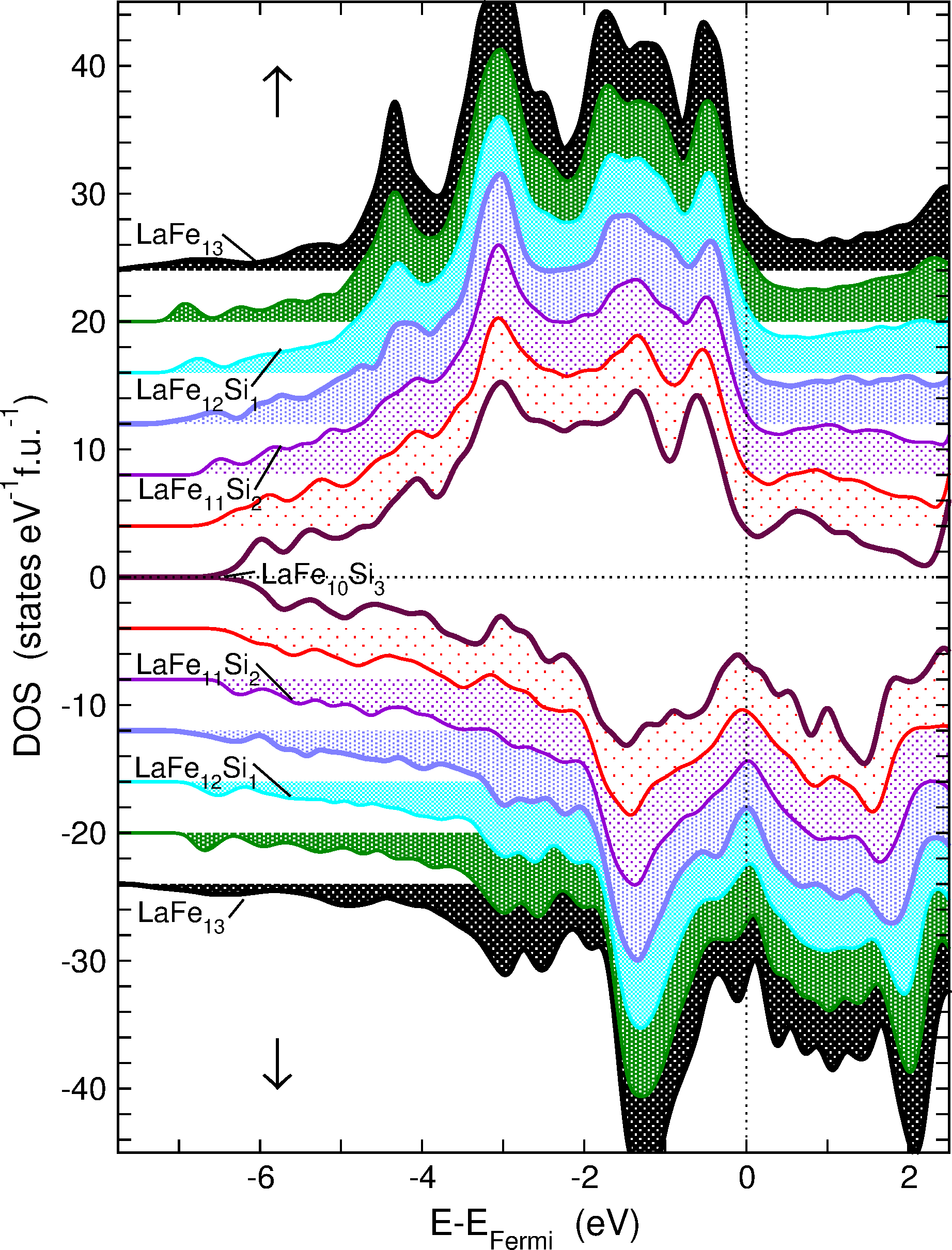}
\caption{%
  Evolution of the total spin-polarized electronic density of states (DOS) of
  LaFe$_{13-x}$Si$_{x}$ as a function of the Si concentration between $x$$\,=\,$0
  and  $x$$\,=\,$3 in steps of $x$$\,=\,$0.5. The Si atoms were placed randomly on the
  (96i) sites using a 112 atom unit cell.
  Again, positive values refer to the majority spin channel,
  negative values to the minority spin channel.
  For better comparison, the DOS curves are shifted by a constant offset
   away from the abscissa with increasing Fe content.
}
\label{fig:SiDOS}
\end{figure}
\begin{figure}[t]%
  \centering
\includegraphics*[width=0.75\linewidth]{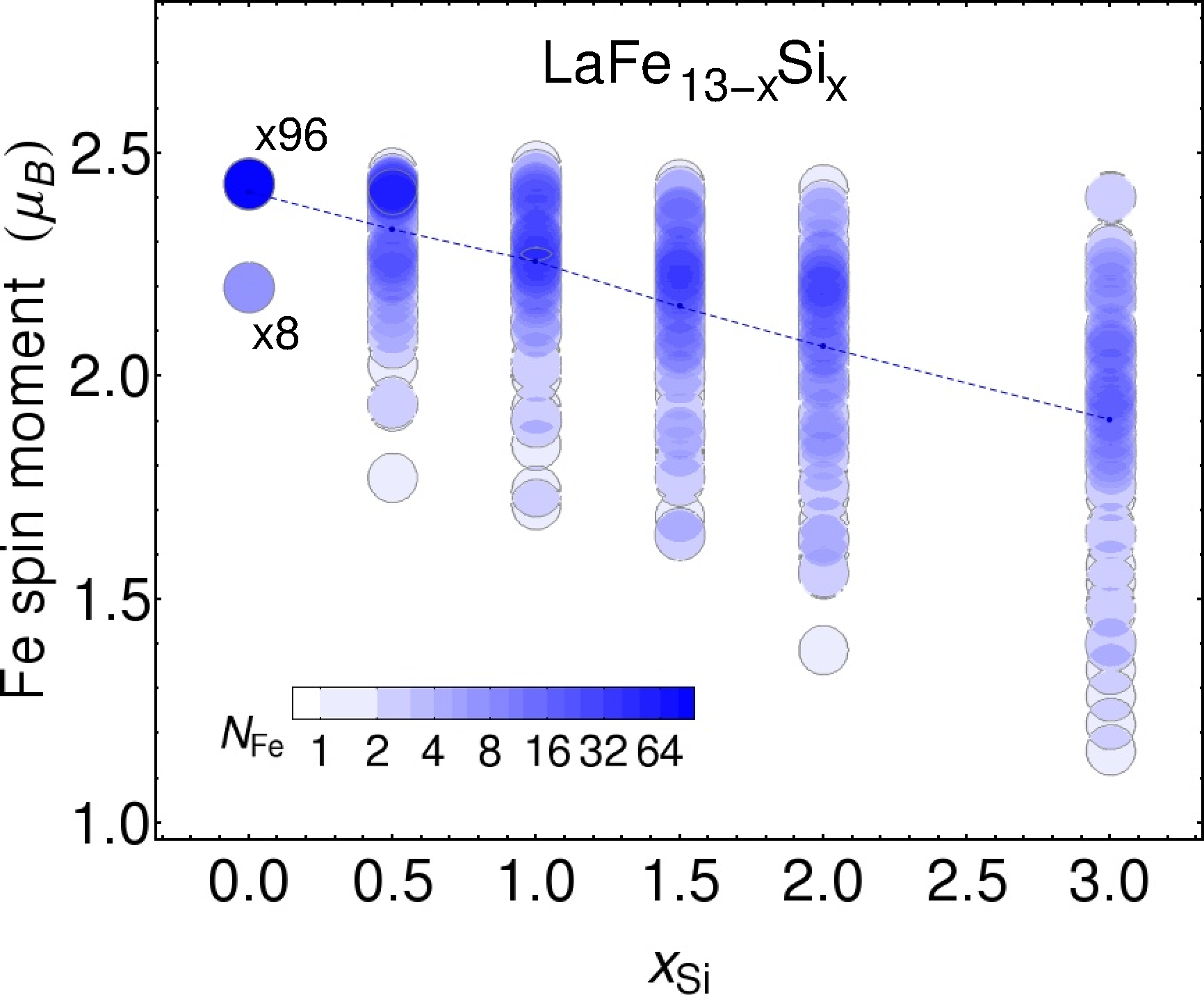}
\caption{%
Site-resolved Fe moments
as a function of Si content $x$ per formula unit
obtained for 112 atom unit cell with random placements of Si.
Blue circles refer to both, Fe$_{\rm I}$ and Fe$_{\rm II}$ sites. Again,
the color becomes more saturated,
when the symbols overlap.
The line denotes the average of all Fe moments for a given composition.
}
\label{fig:SiMoments}
\end{figure}

The question that arises in this context is, whether we can relate the changes in the functional properties
upon variation of the Si content to respective changes of the characteristic minimum in the minority spin DOS
at $E_{\rm Fermi}$, which we identified to be at the heart of the itinerant metamagnetic transition of La-Fe-Si,
leading to its excellent magnetocaloric properties. To test this, we set up a series of
simulations where we varied systematically the amount of Si in the 112 atom unit cell
from $x=0$ to $x=3$ in steps of 0.5.
We restricted our calculations to the ferromagnetic ordered phase, but lattice parameters and interatomic
spacing were subject to atomic relaxations. From these calculations we can derive the evolution
of the electronic density of states (Fig.\ \ref{fig:SiDOS}) and the site-specific magnetic moments of the Fe-sites
(Fig.\ \ref{fig:SiMoments}) as a function of composition. Both confirm indeed our expectations.

The characteristic minimum in the minority spin density of states is most pronounced around $x=1.5$.
With increasing Si content, the average
Fe-moments decrease significantly.
This already motivates the smaller volume change
at the transition arising from the itinerant character of the Fe-moments
as the volume magnetostriction is expected to be proportional to the square of the change in
average magnetic moment.
Another important consequence is the decrease in the local exchange splitting at each site,
which shifts the minimum in the minority spin DOS to lower energies, increasing the number of states
at $E_{\rm Fermi}$.
At the same time the variation in the magnetic moments of the Fe-atoms
becomes larger, resulting from the chemical disorder among their neighbors.
In consequence, the features in the electronic DOS become washed out around the
Fermi level which is weakening the metamagnetic instability behind the first-order transition.
In turn, if we decrease the Si-content, we see that the
distribution of Fe-moments becomes narrower and the features in the electronic DOS become more pronounced.
In addition, a minority spin peak detaches from the large peak at -1.5\,eV, which becomes larger
and moves towards the Fermi energy.
This potentially contributes to the decomposition of the pure LaFe$_{13}$ phase,
since a large number of electronic states close to the highest occupied level
is usually deemed an unstable situation because a redistribution of states, e.\,g., through
structural or chemical rearrangements, will likely benefit from a lower band energy.

\section{Conclusion}
La-Fe-Si owes its excellent magnetocaloric  properties to the partially itinerant, partially localized
nature of the Fe-moments. These vary for a Si-content of $x$$\,=\,$1.5
between approximately $2.2\,\mu_{\rm B}$ in the FM phase
and $1.8\,\mu_{\rm B}$ on average in the PM phase, whereas 
low spin moments $<1.5\,\mu_{\rm B}$ or even a complete quenching are encountered occasionally.
Of particular relevance for the thermodynamic behavior
is here the coupling between the magnitude of the magnetic moment and atomic volume.

Our fixed spin moment calculations which model a ferromagnetic compound with artificially constrained
magnetization prove, that the essential changes in the vibrational density of states observed earlier
at the FM-PM transition, such as the anomalous softening of the phonons leading to the
larger lattice entropy in the PM phase \cite{cn:Gruner15PRL} in combination with the disappearance of a marked
peak at 27\,meV can be solely explained by the reduction of the Fe spin moment, without involving
magnetic disorder. This picture is strongly corroborated by the close agreement between the partial Fe-VDOS obtained
from our FSM calculations and a NRIXS measurement carried out in the PM phase, closely above the Curie temperature.
This means, that future empirical modelling approaches should describe the relevant coupling phenomena
between magnetism and lattice, which determine the transition, in terms of a appropriately designed
{\em moment-volume interaction} -- as opposed to a {\em spin-lattice coupling}, which directly couples
localized spins with the lattice degrees of freedom (e.g., in terms of distance-dependent exchange parameters).
Still, temperature induced magnetic disorder should be regarded as the driving force of
the first-order transition and the large entropy change.

Based on our results, we propose the following scenario: In the FM phase the Fermi level is pinned in a
pronounced minimum of the minority spin density states. This stabilizes a larger spin moment in
the FM phase by the associated gain in band energy
and allows for the competition between two magnetic states at
different volumes. These are
separated by a free-energy barrier which constitutes the first-order nature of the transition and
introduces thermal hysteresis.
Important for the position of the minority spin minimum is the hybridization of the Fe states with the
neighboring atoms. Increasing spin disorder (with increasing temperature) or chemical disorder
(with increasing Si content) weakens this stabilizing feature causing the higher spin moment
and leads thus a to reduced average Fe moment.
The smaller Fe moment with increasing Si content can also
be expected to be the cause of the decreasing volume change.
This explains, why the character of the transition becomes second-order in the end.

In turn, reducing thermal hysteresis by addressing the volume change
likely compromises the intrinsic magnetocaloric properties of the material. 
This applies to the obvious strategy of
shifting the system closer to a second-order transition by increasing the Si content, but also by deliberately introducing magnetic disorder, e.g.,
through an antiferromagnetically coupling element such as Mn. Both approaches reduce the average moment per
Fe-atom in the FM phase. Due to the inherent moment-volume coupling, decreasing the change in average moment per Fe
across the transition also decreases the volume change, which is beneficial in terms of reducing thermal hysteresis.
But due to the very same mechanism this takes at the same time effect on the magnetic, lattice and electronic
entropy change, which contribute cooperatively to the good magnetocaloric properties in the material.
Thus fine tuning of composition and annealing procedures to find a rather homogeneous distribution of atoms
minimizing the variation of the Fe-moments in combination with a microstructure which can accommodate the
volume change at the phase transition might be important prerequisites
regarding the optimization of La-Fe-Si as a magnetocaloric refrigerant.

\begin{acknowledgements}
  We are grateful to Ulrich von H\"orsten (Duisburg-Essen) for valuable technical assistance.
  This work has been supported by the Deutsche Forschungsgemeinschaft in
  the framework of the priority program SPP 1599 (GR3498/3-2, GU514/6-2, WE2623/12-2),
  SPP 1681 (WE2623/7-1), FOR 1509 (WE2623/13-2),
  and by Stiftung Mercator (MERCUR).
  Use of the Advanced Photon Source, an Office of Science User Facility
  operated for the U.S.\ Department of Energy (DOE) Office of Science by
  Argonne National Laboratory, was supported by the U.S.\ DOE
  No.\ DE-AC02-05CH11231).
  The calculations were carried out on the Cray XT6/m and Opterox supercomputer systems
  of the University of Duisburg-Essen. 
\end{acknowledgements}


\providecommand{\WileyBibTextsc}{}
\let\textsc\WileyBibTextsc
\providecommand{\othercit}{}
\providecommand{\jr}[1]{#1}
\providecommand{\etal}{~et~al.}

\end{document}